\documentclass{svmult}

\usepackage{makeidx}         % allows index generation

\usepackage{graphicx}        % standard LaTeX graphics tool
\usepackage{multicol}        % used for the two-column index

\usepackage[bottom]{footmisc}% places footnotes at page bottom

\makeindex             % used for the subject index
\usepackage{units}

\usepackage{units} %\unit[3]{m/s^2} => 3 m/s^2 etc!

\providecommand{\bc}{\begin{center}}
\providecommand{\ec}{\end{center}}
\providecommand{\be}{\begin{equation}}
\providecommand{\ee}{\end{equation}}
\providecommand{\bea}{\begin{eqnarray}}
\providecommand{\eea}{\end{eqnarray}}
\providecommand{\bdm}{\begin{displaymath}}
\providecommand{\edm}{\end{displaymath}}
\providecommand{\bdma}{\begin{eqnarray*}}
\providecommand{\edma}{\end{eqnarray*}}
\providecommand{\ba}{\begin{eqnarray*}}
\providecommand{\ea}{\end{eqnarray*}}
\providecommand{\bi}{\begin{itemize}}
\providecommand{\ei}{\end{itemize}}
\providecommand{\benum}{\begin{enumerate}}
\providecommand{\eenum}{\end{enumerate}}

\providecommand{\refkl}[1]{(\ref{#1})}

\providecommand{\twoCases}[4]{
  \left\{ 
    \begin{array}{ll} 
      #1 & #2 \\
      #3 & #4 
    \end{array} 
  \right.
}

\providecommand{\text}[1]{{\mbox{ #1}}}

\providecommand{\fig}[2]{
   \begin{center}
     \includegraphics[width=#1]{#2}
   \end{center}
}

\providecommand{\abl}[2]{\frac{{\rm d} #1}{{\rm d} #2}}  %d(#1)/d(#2)

\providecommand{\sub}[1]{_{\rm #1}}
\renewcommand{\sup}[1]{^{\rm #1}}

\begin{document}

\title*{From Drivers to Athletes -- Modeling and Simulating
 Cross-Country Sking Marathons}


\author{Martin Treiber\inst{1}\and
Ralph Germ\inst{1}\and
Arne Kesting\inst{2}
}


\institute{Technische Universit\"at Dresden, %W\"urzburger Str. 35, 
01\,062 Dresden, Germany
\texttt{treiber@vwi.tu-dresden.de}
\and TomTom Development Germany GmbH, An den Treptowers 1, 12\,435 Berlin 
\texttt{mail@akesting.de}
}

\maketitle

% Key words: 
% crowd flow
% driven particles
% microscopic traffic flow model for cross-country skiers
% mass sports events
% Vasaloppet
% ski Marathon
% Multi-Model Open-Source Vehicular Simulator MovSim.org

%\vspace{1ex}

\begin{abstract}
 Traffic flow of athletes in classic-style cross-country ski
marathons, with the Swedish \emph{Vasaloppet} as prominent example,
represents a non-vehicular system of 
driven particles with many properties of vehicular traffic flow such
as unidirectional movement, the existence of lanes, and, moreover,
severe traffic jams.
We propose a
microscopic acceleration and track-changing model 
taking into account different fitness levels,
gradients, and interactions between the athletes in all traffic
situations. The model is calibrated on 
microscopic data of the \emph{Vasaloppet 2012}
Using the multi-model open-source simulator {\tt MovSim.org}, we simulate
all 15\,000 participants of the Vasaloppet during the first ten
kilometers.
\end{abstract}

%##########################################################
\section{Introduction}
%##########################################################

\begin{figure}
\fig{0.8\textwidth}{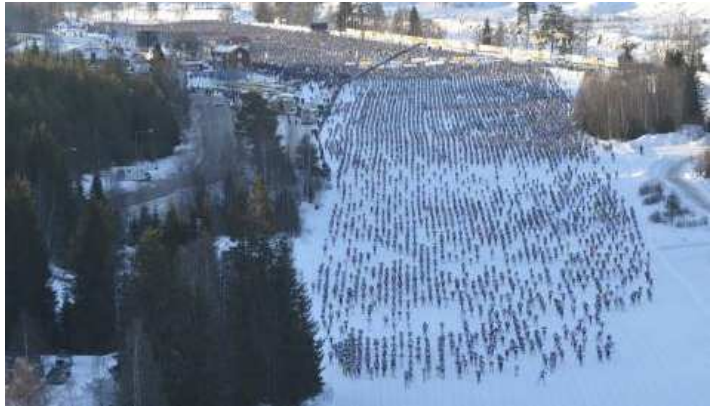}
\caption{\label{fig:photo}Starting phase of the Vasaloppet 2012.}
\end{figure}

Traffic jams are not only observed in vehicular traffic but also
in the crowd dynamics of mass-sport events, particularly cross-country
ski marathons. The Swedish
\emph{Vasaloppet}, a 90-km race  with about 15\,000
participants, is the most prominent example (cf. Fig.~\ref{fig:photo}).
Several other races attract  up to 10\,000
participants. Consequently, ``traffic jams'' among the athletes 
occur regularly. They are not only a hassle for the
athletes but also pose organisational or
even safety threats. 
While there are a few scientific investigations of the
traffic around such events~\cite{ahmadi2011analysis}, 
we are not aware of any investigations on
the crowd dynamics of the skiers
\emph{themselves}.

Unlike the athletes in running or skating events~\cite{TGF13-running}, the skiers in
Marathons for the classic style (which is required in the
Vasaloppet main race) move along fixed tracks, i.e., the traffic flow
is not only unidirectional but \emph{lane based}. 
This allows us to generalize car-following and
lane changing models~\cite{TreiberKesting-Book} to formulate a microscopic model for the
motion of skiers. 

Simulating  the model allows event managers to improve the race
organization by identifying (and possibly eliminating) bottlenecks, determining the optimum
number of starting groups and 
the maximum size of each
group, or optimizing the starting schedule~\cite{TGF13-running}.
%, e.g., going from the
%traditional mass start to a ``wave start'' where there is a time delay
%between the start of each starting group. 

We propose a
microscopic acceleration and track-changing model 
for cross-country skiers taking into account different fitness levels,
gradients, and interactions between the athletes in all traffic
situations. After calibrating the model on 
microscopic data of jam free sections of the \emph{Vasaloppet 2012},
we apply the open-source 
simulator {\tt MovSim.org}~\cite{movsim} to simulate 
all 15\,000 participants of the Vasaloppet during the first ten
kilometers. The simulations show that the initial jam causes a delay of
up to \unit[40]{minutes} which agrees with evidence from the data.

The next section introduces the model. In Section~\ref{sec:sim}, we
describe the calibration, the simulation, and the
results. Section~\ref{sec:concl} concludes with a discussion.

%#########################################
%#########################################

%##########################################################
\section{\label{sec:mod}The Model}
%##########################################################

\begin{figure}
\fig{0.98\textwidth}{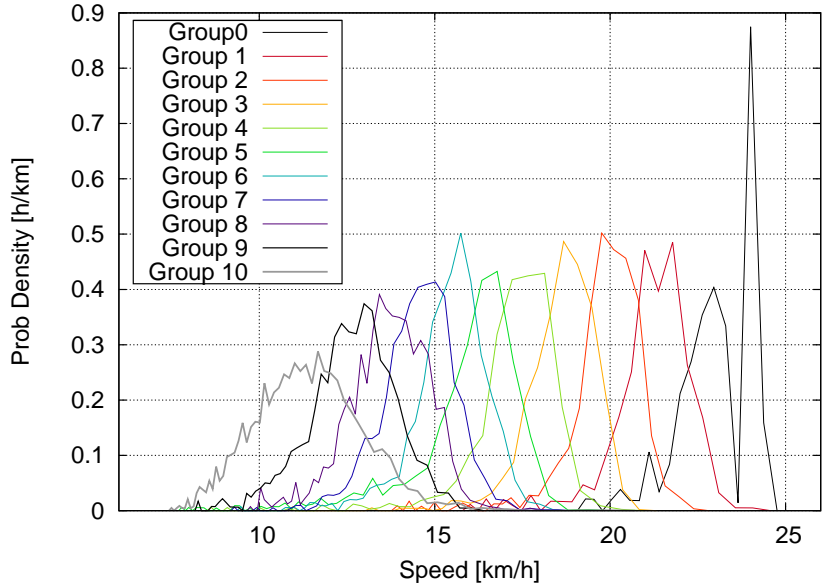}
\caption{\label{fig:speeds}Speed density functions for the section
  between Station~1 and~2 for each starting group. No jams were
  observed in this section.}
\end{figure}

Unlike the normal case in motorized traffic,
the ``desired'' speed (and acceleration) of a skier is
restricted essentially by his or her performance
(maximum mechanical power $P=P\sub{max}$),
and by the maximum speed $v_c$ for active propulsion ($P=0$ for
$v \ge v_c$). Since, additionally,
 $P\to
0$ for $v\to 
0$, it is plausible to model the usable power as a function of
the speed as a parabola, 
\be
\label{pow1}
P(v,v_c)=4 P\sub{max}\frac{v}{v_c}\left(1-\frac{v}{v_c}\right) \theta(v_c-v),
\ee
where $\theta(x)=1$ if $x\ge 0$, and zero, otherwise.
While the maximum mechanical power is reached at $v_c/2$, the maximum
propulsion force $F\sub{max}=4P\sub{max}/v_c$, and the maximum acceleration
\be
a\sub{max}=\frac{4P\sub{max}}{mv_c},
\ee
is reached at zero speed.
The above formulas are valid for conventional techniques such as the
``diagonal step'' or ``double poling''. However, if the uphill
gradient (in radian) exceeds the angle $\alpha\sub{slip}=a\sub{max}/g$
(where $g=\unit[9.81]{m/s^2}$), no forward movement is possible in this
way. Instead, when $\alpha>\alpha\sub{max}/2$, athletes use the slow
but steady ``fishbone'' 
style described by~\refkl{pow1} with a lower maximum speed $V_{c2}$
corresponding to a higher  maximum gradient $4P\sub{max}/(gmv_{c_2})$. In summary, the
propulsion force reads
\be
F(v,\alpha)=\twoCases
{P(v,v_c)/v}{\alpha<\alpha\sub{max}/2}
{P(v,v_{c2})/v}{\text{otherwise.}}
\ee
Balancing this force with the inertial, friction, air-drag, and
gravitational 
forces defines the free-flow acceleration $\dot{v}\sub{free}$:
\be
m\dot{v}\sub{free} =F(v) - \frac{1}{2}c_d A \rho v^2
- mg(\mu_0+\alpha).
\ee
If the considered skier is following a leading athlete (speed $v_l$) at
a spatial gap $s$, the free-flow acceleration is complemented by the
decelerating interaction force of the intelligent-driver model
(IDM)\cite{TreiberKesting-Book} leading to the full longitudinal model
\be
\abl{v}{t}=\text{min}\left\{\dot{v}\sub{free},\,
  a\sub{max}\left[1-\left(\frac{s^*(v,v_l)}{s}\right)^2\right]\right\},
\ee
where the desired dynamical gap of the IDM depends on the gap $s$ and the
leading speed $v_l$ according to
\be
s^*(v,v_l)=s_0+\max\left(0,\, vT+\frac{v (v_l-v)}
   {2\sqrt{a\sub{max}b}}\right). 
\ee
Besides the ski length, this model has the parameters $c_d A \rho/m$,
$\mu_0$, $P\sub{max}/m$, $v_c$ 
(defining $a\sub{max}$),
$v_{c2}$, $s_0$, $T$, and $b$ (see Table~\ref{tab:param}). It is
calibrated such that the maximum unobstructed speed $v\sub{max}$ on
level terrain, 
defined by $F(v\sub{max},0)-c_d A \rho v\sub{max}^2/2-
mg\mu_0=0$, satisfies the observed speed distributions on level
unobstructed sections (Fig.~\ref{fig:speeds}).

\subsection{Lane-changing model}
We apply the general-purpose  lane−changing model
MOBIL~\cite{TreiberKesting-Book}. Generally, lane changing and
overtaking is allowed on either side and 
crashes are much less avoided than in vehicular traffic, so, the
symmetric variant of the model
with zero politeness and rather aggressive safety settings is appropriate. Lane changing
takes place if it is both safe and advantageous. The safety criterion
is satisfied if, as a consequence of the change, 
 the back skier on the new track is not forced to decelerate by
more than his or her normal deceleration ability $b$:

\be
\label{safety}
\abl{v\sub{back,new}}{t} \ge -b.
\ee
A change is advantageous if, on the new track, the athlete can
accelerate more (or needs to decelerate less) than on the old track:
\be
\abl{v\sub{front,new}}{t} \ge \abl{v\sub{actual}}{t} + \Delta a,
\ee
where the only new parameter $\Delta a$ represents
 some small threshold to avoid lane changing for
marginal advantages. Note that for mandatory lane changes (e.g., when a track
ends), only the safety criterion~\refkl{safety} must be satisfied.

%#########################################
%#########################################

%########################################
\begin{table}
\centering
\caption{\label{tab:param}Model parameters of the proposed longitudinal model}
\begin{tabular}{ll}
\hline\noalign{\smallskip}
Parameter & Typical Value ($4\sup{th}$ starting group)\\
\noalign{\smallskip}\hline\noalign{\smallskip}
ski length $l$ & \unit[2]{m} \\
Mass $m$ incl. equipment & \unit[80]{kg} \\
air-drag coefficient $c_d$ & 0.7 \\
frontal cross section $A$  & $\unit[1]{m^2}$ \\ 
friction coefficient $\mu_0$ & 0.02 \\ \hline
maximum mechanical power $P\sub{max}$ & \unit[150]{W}\\
limit speed for active action $v_c$ & \unit[6]{m/s} \\
time gap $T$ & \unit[0.3]{s} \\
minimum spatial gap $s_0$ & \unit[0.3]{m} \\
normal braking deceleration $b$ & $\unit[1]{m/s^2}$ \\
maximum deceleration  $b\sub{max}$ & $\unit[2]{m/s^2}$ \\
\noalign{\smallskip}\hline
\end{tabular}
\end{table}
%########################################

%##########################################################
\section{\label{sec:sim}Simulation Results}
\begin{figure}
\fig{0.8\textwidth}{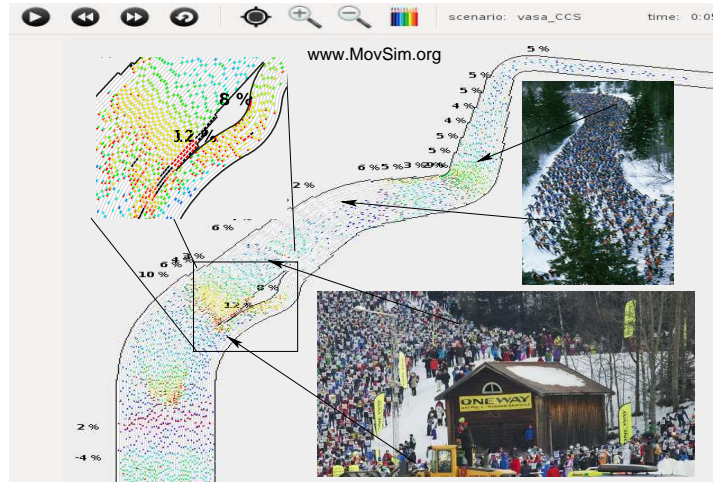}
\caption{\label{fig:sim}Screenshot of the MovSim Simulation of the
  first \unit[10]{km} of the Vasaloppet~2012 (center) with an
  enlargement of the diverge-merge section (left top). Also shown are
  two photos of the crowd flow at
  the corresponding sections (right).}
\end{figure}

%##########################################################
We have simulated all of the 15\,000 athletes of
the Vasaloppet~2012 for the first \unit[10]{km}
(cf. Fig.~\ref{fig:sim}) by implementing the model into the
open-source traffic simulator {\tt MovSim.org}. The starting field 
includes 70 parallel 
tracks (cf. Fig.~\ref{fig:photo}) where the 10 starting groups (plus a small elite group) are
arranged in order. Further ahead, the number of tracks decreases
gradually down to 8~tracks at the end of the uphill section for $x\ge
\unit[7]{km}$. The uphill gradients and the course geometry (cf. Fig.~\ref{fig:sim}) were
obtained using Google Earth.

%#########################################
%#########################################

%#########################################
\begin{figure}
\fig{0.98\textwidth}{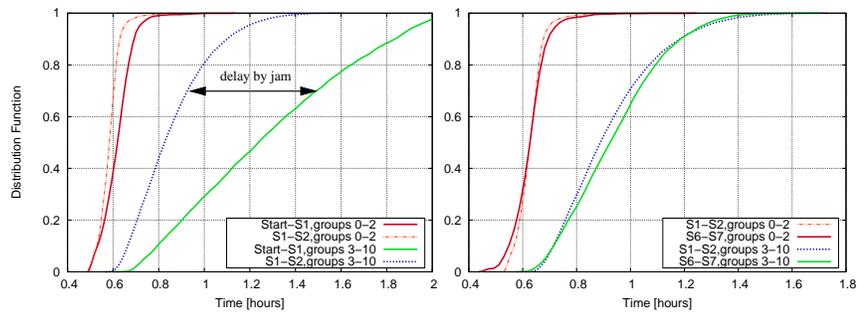}
\caption{\label{fig:tdistr}Distribution functions of the split times
  from the start to Station S1 (left), S1 to S2 (left and right), and
  S6 to S7 (right), shown separately for the fastest groups (elite and
  groups 1 and 2) and the remaining
  groups~3 to~10. All three sections take about the same time. 
Major jams occur only for the groups~3 to~10 and
  only between the start and S1.}
\end{figure}
%#########################################

As in the real event, we simulated a mass start.
 While the initial starting configuration dissolves
relatively quickly, massive jams form at the beginning of the gradient
section, particularly at the route divide (inset of
Fig.~\ref{fig:sim}).
In summary, the delays due to the jams accumulated  up to \unit[40]{minutes}
for the last starting groups which agrees with the macroscopic
flow-based analysis of the split-time data (Fig.~\ref{fig:tdistr}).

%We have also simulated a delayed wave-start scheme. A delay of
%just \unit[5]{minutes} between each starting group essentially
%eliminated the congestion for all but the last starting group.

%##########################################################
\section{\label{sec:concl}Conclusion}
%##########################################################

Using the open-software MovSim, we have quatitatively  reproduced the
congestions and stop-and-go waves on the first ten kilometers  of the Vasaloppet
Race 2012. The jams leading to a delay of up to 40~minutes are caused a steep
uphill section and a simultaneous reduction of the number of
tracks. Further simulations have also shown that eliminating the worst
bottlenecks by locally adding a few tracks only transfers the jams to
locations further downstream. In contrast, replacing the mass start
(which is highly controversial) by
a wave start with a
five-minute delay between the starting groups would essentially
eliminate the jams without the need to reduce the total number of
participants.

%In summary, we have shown that a dynamic multi-particle simulation using a novel
%microscopic model can help
%optimize  large-skale skiing events by anticipating
%the effects of new starting schemes, modificatiuons of the course, or
% limitations of the number of
%participants. 
%It would be interesting to generalize this research to
%non-lane-based traffic flow including running and skating events, but
%also mixed vehicular flow in general.

%##########################################################

\bibliographystyle{elsart-num}

\bibliography{database,databaseSpecialTGF13}

\end{document}